# Citation advantage of Open Access articles likely explained by quality differential and media effects

In a study of articles published in the *Proceedings of the National Academy of Sciences*, Gunther Eysenbach discovered a significant citation advantage for those articles made freely-available upon publication (Eysenbach 2006). While the author attempted to control for confounding factors that may have explained the citation differential, the study was unable to control for characteristics of the article that may have led some authors to pay the additional page charges ($1,000) for immediate OA status. For instance, of the 1,492 original articles (212 OA) tracked in Eysenbach's study, OA articles were more than twice as likely to be featured on the front cover of the journal (3.3% vs. 1.4%).

Coverage in the popular press is well known to amplify the transmission of scientific information to the research community. As a result, articles that receive coverage in newspapers are more likely to be cited (Phillips et al. 1991) (Kiernan 2003). Of the articles studied in Eysenbach's cohort, OA articles were nearly twice as likely to be picked up by the media (15% vs. 8%) and when cited reached, on average, nearly twice as many news outlets as subscription-based articles (4.2 vs. 2.6).

In order to do publish one's results in a prestigious peer-reviewed journal indexed by ISI's Web of Science (the tool by which Eysenbach counts citations), one is likely to be located at an institution with both access to resources for doing research as well as adequate access to the literature. Considering that PNAS has one of the highest circulation rates of any scientific journals, that it provides free and immediate access to readers in developing countries, that it employs a tiered pricing model that allows small institutions to pay a fraction of the cost as large institutions, and that sharing electronic copies of articles is easier than ever, it is difficult to believe that immediate access in a journal that provides free access to all of its articles after six months would result in a citation advantage.

The fact that OA articles were more likely to be featured on the front cover of PNAS and covered by the media suggests that other causal explanations may explain the OA advantage. Open Access may be a result – not a cause of – a quality differential which is amplified by the media. While Eysenbach's attempt to control other explanatory variables was excellent, what is needed are true randomized controlled studies of OA publishing.

**Research Notes:**

Major Papers, Magazines and Journals categories were searched in Lexis/Nexis between May 1, 2004 and Jan 31, 2005 for media coverage of articles published in PNAS between June 8 and Dec 28, 2004. The additional period was to cover articles published online before print. The news sections of the following journals were also searched since they regularly contain coverage of articles published in PNAS: *Science*, *Nature*, the *New Scientist*, *Journal of the American Medical Association*, *BMJ*, and *Scientific American*. Lastly, as additions to general news sources, *The Chronicle of Higher Education*, The Associated Press, and NPR transcripts were also searched.

**References:**

Eysenbach G (2006) Citation Advantage of Open Access Articles. *PLoS Biology* 4(5).

Kiernan V (2003) Diffusion of news about research. *Science Communication* 25(1): 3-13.

Phillips D, Kanter E, Bednarczyk B, Tastad P (1991) Importance of the lay press in the transmission of medical knowledge to the scientific community. *New England Journal of Medicine* 325(16): 1180-1183.